# Interpretation and visualization of distance covariance through additive decomposition of correlations formula

Andi Wang[1], Hao Yan[1], and Juan Du[2]

1. Arizona State University
2. Hong Kong University of Science and Technology

**Abstract**

Distance covariance is a widely used statistical methodology for testing the dependency between two groups of variables. Despite the appealing properties of consistency and superior testing power, the testing results of distance covariance are often hard to be interpreted. This paper presents an elementary interpretation of the mechanism of distance covariance through an additive decomposition of correlations formula. Based on this formula, a visualization method is developed to provide practitioners with a more intuitive explanation of the distance covariance score.

## 1. Introduction

Testing for the independence between two types of measurements has been a basic problem in statistics and has significant importance in engineering applications. Among the literature on independence testing methods, the distance covariance (Lyons 2013, Székely, et al. 2007) enjoys an appealing property of being statistically consistent against all forms of dependencies and has been recognized for its high testing power (Sarkar and Ghosh 2018, Simon and Tibshirani 2014). Since distance covariance has been proposed, the method has been extended and adapted for multiple purposes, including the feature screening (Li, et al. 2012) and causal inference (Chakraborty and Zhang 2021). It is also extended for multiple types of data (Zhou 2012). On another line of dependence testing approach, Gretton, et al. (2005), Gretton, et al. (2005), Gretton, et al. (2008), and Gretton, et al. (2012) proposed several kernel-based methods for measuring independence. Distance covariance is shown to have an equivalent relationship with Hilbert Schmidt Independence Criterion (HSIC), one of the kernel-based dependence measures (Sejdinovic, et al. 2013) when the kernel functions of HSIC and the distance metrics of distance covariance are selected correspondently.

Despite the strong testing power, the lack of interpretation remains a critical barrier to the widespread application of distance covariance in the engineering field. Given the measurements from two groups of variables, distance covariance determines whether there is strong evidence to reject the independent hypothesis between these two groups of variables. However, even when the independent hypothesis is rejected, practitioners still do not know how and why the variables are related. This lack of interpretation limits its application in engineering. For example, practitioners in the manufacturing industry often aim to investigate the relationship between manufacturing process variables and product quality variables. Upon the rejection of the independent hypothesis, distance covariance fails to provide further insights into the relationship between the two groups of variables, impeding the diagnostic or modeling process and hindering the achievement of quality improvement goals.

Although there are multiple understandings of the distance covariance test statistics, none of them provide an explicit and intuitive understanding of the relationship between variables. In the original version of distance covariance (Székely, Rizzo and Bakirov 2007), the distance covariance is interpreted as a weighted $L_2$ norm of $f_{XY} - f_X f_Y$, where $f_{XY}$, $f_X$ and $f_Y$ are characteristic functions of the joint distribution $(X, Y)$, and the marginal distributions $X$ and $Y$. The proposition 3.7 of Lyons (2013) associated the distance covariance with the norm of a barycenter map. As for HSIC, the population statistic is defined as the norm of the cross-covariance operator. However, all the above interpretations are not intuitive and requires abstract concepts of probability theory and functional analysis, making it difficult to develop illustrations and visualization approaches for the relationship between two groups of variables, especially for high-dimensional data.

In this paper, we propose an elementary interpretation for distance covariance. This interpretation is based on a key result of the paper: the formula of *additive decomposition of correlations* (ADC), which explicitly shows that *distance covariance between X and Y is the weighted sum of the correlations between all pairs of features $\phi_i(X)$ and $\psi_j(Y)$, where $\{\phi_i(X)\}$ and $\{\psi_j(Y)\}$ are sets of features generated from X and Y*. We further established the connection between the sample statistics of HSIC and distance covariance: the data $\{\mathbf{x}_i\}_{i=1}^n$ and $\{\mathbf{y}_i\}_{j=1}^n$ are transformed to $n$ orthogonal feature vectors, and the sample statistic is a weighted sum of $n^2$ pairs of features among them. Based on these results, we propose a data visualization method that displays both sets of features generated from $X$ and $Y$ and the correlations between them. The practitioner can evaluate the engineering implications of the features to decide whether the selection of semi-metrics that defines the distance covariance are appropriate. The visualization also enables practitioners to understand what leads to the rejection of the independent hypothesis, thereby leading to a clearer interpretation of the test result and enabling the practitioners, especially non-statisticians in engineering sectors to confidently utilize the distance covariance method.

The remaining part of the article will be organized as follows. In Section 2, we provide an overview of the relevant background knowledge and existing works related to our interpretation. In Section 3, we present the main results of ADC formula, which forms the foundation of our interpretation of distance covariance. We will then describe the visualization method developed using the ADC formula. In Section 4, we demonstrate the using the visualization method through several examples. We use the visualization approach on a dataset of solar cell manufacturing in Section 5. Finally, Section 6 concludes the article.

## 2. Overview of distance covariance and Hilbert-Schmidt independent criterion

This section reviews the distance covariance method (Lyons 2013, Székely, Rizzo and Bakirov 2007), the HSIC (Gretton, Borgwardt, Rasch, Schölkopf and Smola 2012, Gretton, Bousquet, Smola and Scholkopf 2005). Specifically, we will highlight the existing results on the association between distance covariance and HSIC (Sejdinovic, Sriperumbudur, Gretton and Fukumizu 2013), which is the prerequisite for understanding the ADC formula.

### 2.1. *Distance covariance for testing statistical independence.*

Consider a standard setup where two random vectors $X \in \mathcal{X} = \mathbb{R}^p$ and $Y \in \mathcal{Y} = \mathbb{R}^q$ represent the data collected from two groups of variables. The joint distribution of $(X, Y)$ is $P_{XY}$. In this article, the distance covariance we discuss follows the *generalized distance covariance* introduced in Lyons (2013), while we drop the word "generalized" thereafter for simplicity. Let $d(\cdot,\cdot)$ and $\rho(\cdot,\cdot)$ be two semi-metrics of negative types defined on $\mathcal{X}$ and $\mathcal{Y}$ that satisfy $\mathbb{E}[d^2(x_0, X)] < \infty$ and $\mathbb{E}[\rho^2(y_0, Y)] < \infty$ for some $x_0 \in \mathcal{X}$ and $y_0 \in \mathcal{Y}$, the generalized distance covariance for the population $P_{XY}$ is defined as

$$V(P_{XY}, d, \rho) = \mathbb{E}_{X,Y}\mathbb{E}_{X',Y'}d(X,X')\rho(Y,Y') + \mathbb{E}_X\mathbb{E}_{X'}d(X,X')\mathbb{E}_Y\mathbb{E}_{Y'}\rho(Y,Y')$$
$$- 2\mathbb{E}_{XY}[\mathbb{E}_{X'}d(X,X')\mathbb{E}_{Y'}\rho(Y,Y')],$$

where $(X, Y), (X', Y') \overset{IID}{\sim} P_{XY}$. Given IID sample $\mathcal{D} = (\mathbf{x}_i, \mathbf{y}_i)_{i=1}^n \sim P_{XY}$, the sample statistic for this generalized distance covariance $\hat{V}(\mathcal{D}, d, \rho)$ can be calculated from the following procedure:

1. Calculate the pairwise distance metrics between every two samples of $X$ and $Y$, $\mathbf{D} = d(\mathbf{x}_i, \mathbf{x}_{i'})_{n \times n}$, $\mathbf{R} = \rho(\mathbf{y}_j, \mathbf{y}_{j'})_{n \times n}$.

2. Calculate $\widetilde{\mathbf{D}} = \left(\mathbf{I} - \frac{1}{n}\mathbf{J}\right)\mathbf{D}\left(\mathbf{I} - \frac{1}{n}\mathbf{J}\right)$ and $\widetilde{\mathbf{R}} = \left(\mathbf{I} - \frac{1}{n}\mathbf{J}\right)\mathbf{R}\left(\mathbf{I} - \frac{1}{n}\mathbf{J}\right)$ to center the rows and columns of $\mathbf{D}$ and $\mathbf{R}$, where $\mathbf{J} = \mathbf{1}_n\mathbf{1}_n^\top$ is an $n \times n$ matrix of 1s and $\mathbf{I}$ is the order-$n$ identity matrix.

3. $\widehat{V}(\mathcal{D}, d, \rho) = \frac{1}{n^2}\left[\text{vec}\,\widetilde{\mathbf{D}}\right]^\top\left[\text{vec}\,\widetilde{\mathbf{R}}\right] = \frac{1}{n^2}\text{tr}(\widetilde{\mathbf{D}}^\top\widetilde{\mathbf{R}})$.

Both population and sample distance covariance are non-negative. The sample statistics $\widehat{V}(\mathcal{D}, d, \rho)$ gives a consistent estimation of the population counter part $V(P_{XY}, d, \rho)$. Lyons (2013) shows that if the distance metrics $d$ and $\rho$ are both strong negative types, $V(P_{XY}, d, \rho) = 0$ only when $X$ and $Y$ are independent. Therefore, the hypothesis testing procedure can be defined based on the asymptotic distribution of $\widehat{V}(\mathcal{D}, d, \rho)$ (Theorem 2.7 of Lyons (2013)), with its rejection region being $\widehat{V}(\mathcal{D}, d, \rho) > H$, where the threshold $H$ is defined by the prescribed type I error rate.

## 2.2. HSIC and its relationship with distance covariance

The Hilbert-Schmidt independence criterion is an independency testing procedure defined based on the reproducing kernel Hilbert spaces (Aronszajn 1950) and distribution embedding. Let $k(\cdot,\cdot)$ and $l(\cdot,\cdot)$ be symmetric, positive definite kernel functions on $\mathcal{X} \times \mathcal{X}$ and $\mathcal{Y} \times \mathcal{Y}$ that satisfy $\mathbb{E}_X[k(X,X)] < \infty$ and $\mathbb{E}_X[l(Y,Y)] < \infty$ respectively. Gretton, et al. (2005) proposed HSIC as the squared Hilbert-Schmidt norm of the covariance operator between two RKHS $\mathcal{X}$ and $\mathcal{Y}$ with respective kernels $k$ and $l$, and expressed HSIC$(P_{XY}, k, l)$ in their Lemma 1 with

$$\text{HSIC}(P_{XY}, k, l) = \mathbb{E}_{X,X',Y,Y'}[k(X,X')l(Y,Y')] + \mathbb{E}_{X,X'}k(X,X')\mathbb{E}_{Y,Y'}l(Y,Y') \\ -2\mathbb{E}_{X,Y}[\mathbb{E}_{X'}k(X,X')\mathbb{E}_{Y'}l(Y,Y')], \tag{1}$$

where $(X, Y), (X', Y')$ are two independent samples from the distribution $P_{XY}$. Under certain conditions, HSIC$(P_{XY}, k, l)$ equals to zero if and only if $X$ and $Y$ are independent. The empirical HSIC is given as $\widehat{\text{HSIC}}(\mathcal{D}; k, l) = \frac{1}{n^2}\text{tr}(\mathbf{KHLH})$, where $\mathbf{K} = [k(\mathbf{x}_i, \mathbf{x}_{i'})]_{n \times n}$ and $\mathbf{L} =$ $[k(\mathbf{y}_i, \mathbf{y}_{i'})]_{n \times n}$. The empirical HSIC is also used as a testing statistic for independence because it is a consistent estimation for HSIC $(P_{XY}, k, l)$.

Sejdinovic, Sriperumbudur, Gretton and Fukumizu (2013) established the connection between HSIC and distance covariance under a stricter condition on kernels, $k \in L^2(\mathcal{X}^2, P_X^2)$, $l \in L^2(\mathcal{Y}^2, P_Y^2)$. Let the distance semi-metrics be generated by the kernels through

$$d(x, x') = k(x, x) + k(x', x') - 2k(x, x'); \; \rho(y, y') = l(y, y) + l(y', y') - 2l(y, y'),$$

The distance covariance and the HSIC are related with

$$V(P_{XY}, d, \rho) = 4 \cdot \text{HSIC}(P_{XY}, k, l). \tag{2}$$

Using this connection between HSIC and distance covariance, this article derives an additive decomposition of correlations formula from to interpret the distance covariance.

## 3. Additive decomposition of correlations and the visualization methods.

Recall that both distance covariance or HSIC are consistent against any dependence alternative, while their definition does not reveal an intuitive explanation of their mechanism. In this section, we give an intuitive interpretation of them by presenting an additive decomposition of correlations formula for both population and sample statistics. In short, our interpretation of distance covariance can be described with two key points:

- The distance semi-metrics or the kernel function in respective domains $\mathcal{X}$ and $\mathcal{Y}$ specifies a set of features with weights (importance indices) in either domain through Mercer's decomposition.
- The distance covariance is a weighted-sum-of-squared correlation of all pairs of features.

Based on the formula, we develop a visualization method for distance covariance.

*3.1. ADC formula of population distance covariance*

Under the setup discussed in Section 2, let the marginal distributions of data $(X, Y)$ be $P_X(x) = \int P_{XY}(x, y) dy$ and $P_Y(y) = \int P_{XY}(x, y) dx$ respectively. We will interpret the distance covariance with the HSIC due to the connection (2), and therefore we generate the following kernel functions.

$$\begin{aligned} k(x, x') &= -d(x, x') + \mathbb{E}_{X \sim P_X}[d(X, x')] + \mathbb{E}_{X' \sim P_X}[d(x, X')] - \mathbb{E}_{X \sim P_X}\mathbb{E}_{X' \sim P_X}[d(X, X')] \\ l(y, y') &= -\rho(y, y') + \mathbb{E}_{Y \sim P_Y}[\rho(Y, y')] + \mathbb{E}_{Y' \sim P_Y}[\rho(y, Y')] - \mathbb{E}_{Y \sim P_Y}\mathbb{E}_{Y' \sim P_Y}[\rho(Y, Y')] \end{aligned} \quad (3)$$

The kernels are defined such that they generate the distance metrics $d(\cdot, \cdot)$ and $\rho(\cdot, \cdot)$ respectively, according to Definition 17 of Sejdinovic, Sriperumbudur, Gretton and Fukumizu (2013). Also, this specific form of kernels is centered at marginal distributions $P_X$ and $P_Y$ respectively with $\mathbb{E}_{x, x' \overset{IID}{\sim} P_X}[k(x, x')] = 0$ and $\mathbb{E}_{y, y' \overset{IID}{\sim} P_Y}[l(y, y')] = 0$ (see Equation (4.4) of Sejdinovic, Sriperumbudur, Gretton and Fukumizu (2013)), and we will see later that it leads to features with zero mean.

The key step to deriving the ADC formula is to apply the eigen decomposition for the kernel functions $k(\cdot, \cdot)$ and $l(\cdot, \cdot)$ in the expression of HSIC. By Mercer's theorem of $\sigma$-compact space (Sun 2005), we can express the kernels $k(\cdot, \cdot)$ and $l(\cdot, \cdot)$ using $k(x, x') = \sum_{i=1}^{\infty} \lambda_i \phi_i(x) \phi_i(x')$ and $l(y, y') = \sum_{j=1}^{\infty} \sigma_j \psi_j(y) \psi_j(y')$. Ensured by the centered kernels $k(\cdot, \cdot)$ and $l(\cdot, \cdot)$, $\{\phi_i\}$ and $\{\psi_j\}$ are orthonormal function basis regarding the probability measure $P_X(x)$ and $P_Y(y)$ respectively.

The formulation leads to the following result, the additive decomposition formula of HSIC.

**Proposition 1** Suppose that kernel functions $k \in L^2(\mathcal{X}, P_X)$, $l \in L^2(\mathcal{Y}, P_Y)$ defined in (3) has Mercer decompositions $k(x, x') = \sum_{i=1}^{\infty} \lambda_i \phi_i(x) \phi_i(x')$ and $l(y, y') = \sum_{j=1}^{\infty} \sigma_j \psi_j(y) \psi_j(y')$ regarding distribution $P_X$ and $P_Y$, respectively. The HSIC in Equation (1) can be derived as

$$\text{HSIC}(P_{XY}, k, l) = \sum_{i=1}^{\infty} \sum_{j=1}^{\infty} \lambda_i \sigma_j \big(\text{corr}[\phi_i(X), \psi_j(Y)]\big)^2. \quad (4)$$

The derivation of this proposition is given in Appendix A. Proposition 1 explicitly expresses how HSIC involves the correlation between features implicitly defined by the kernel functions. With Equation (2), we immediately have the additive decomposition formula of the distance covariance:

$$V(P_{XY}; d, \rho) = 4 \sum_{i=1}^{\infty} \sum_{j=1}^{\infty} \lambda_i \sigma_j \left(\text{corr}[\phi_i(X), \psi_j(Y)]\right)^2. \tag{5}$$

Equation (5) provides a convenient interpretation of how distance covariance detects statistical dependence. The eigen functions $\{\phi_i(x)\}$ and $\{\psi_j(y)\}$ can be seen as two sequences of functions that generate infinite features from the functions of $\mathbb{R}^p \to \mathbb{R}$ and $\mathbb{R}^q \to \mathbb{R}$ respectively for evaluation the dependency between $X$ and $Y$. The distance covariance is the sum of their pairwise correlations $\text{corr}[\phi_i(X), \psi_j(Y)]$ weighted by $\lambda_i \sigma_j$'s. For simple distribution and the kernel function $k$ and $l$, we found that the eigen functions $\phi_i, \psi_j$ corresponding to smaller eigen values $\lambda_i$'s and $\sigma_j$'s are more complex features. Therefore, the expression (5) automatically assigns smaller weights for the squared correlation of complex features and assign larger weights for the squared correlation of simpler features. When $X$ and $Y$ are independent, the correlation between any features generated from $X$ and $Y$ are also independent, thereby resulting in zero distance covariance.

### *3.2. ADC formula of sample distance covariance*

Next, we present the corresponding ADC formula for the sample distance covariance. Note that Sejdinovic, Sriperumbudur, Gretton and Fukumizu (2013) only illustrated the relationship between population distance covariance and HSIC, but not the relationship between the *empirical* HSIC and the *sample* distance covariance. We first establish this relationship through pure algebraic derivation.

**Proposition 2** Given data $\mathcal{D} = (\mathbf{x}_i, \mathbf{y}_i)_{i=1}^n$, we have

$$\hat{V}(\mathcal{D}; d, \rho) = 4 \cdot \widehat{\text{HSIC}}(\mathcal{D}; k, l) \qquad (6)$$

if the distance metrics $d$ and $\rho$ are generated by $k$ and $l$ respectively.

The proof is given in Appendix B. Then, Proposition 3 below gives the ADC formula for the sample HSIC.

**Proposition 3** The closed-form expression of the empirical HSIC is given by

$$\widehat{\text{HSIC}}(\mathcal{D}; k, l) = \frac{1}{n^2} \sum_{i=1}^n \sum_{j=1}^n \lambda_i \sigma_j \, \widehat{\text{corr}}(\boldsymbol{\phi}_i, \boldsymbol{\psi}_j)^2, \qquad (7)$$

where $\{\lambda_i, \boldsymbol{\phi}_i\}$ are the eigenvalues and eigenvectors of the matrix $\mathbf{HKH}$, and $\{\sigma_j, \boldsymbol{\psi}_j\}$ are the eigenvalues and eigenvectors of the matrix $\mathbf{HLH}$, where $\mathbf{H} = \mathbf{I} - \frac{1}{n}\mathbf{J}$, $\mathbf{K} = \left(k(\mathbf{x}_i, \mathbf{x}_j)\right)_{n \times n}$, and $\mathbf{L} = \left(l(\mathbf{y}_i, \mathbf{y}_j)\right)_{n \times n}$. $\widehat{\text{corr}}(\cdot, \cdot)$ denotes the sample correlation between two vectors.

The proof of Proposition 3 is in Appendix C. The results of Proposition 2 and Proposition 3 lead to the following ADC formula of sample distance covariance:

$$\hat{V}(\mathcal{D}; d, \rho) = \frac{4}{n^2} \sum_{i=1}^n \sum_{j=1}^n \lambda_i \sigma_j \, \widehat{\text{corr}}(\boldsymbol{\phi}_i, \boldsymbol{\psi}_j)^2.$$

The ADC formula for sample distance covariance share many similarities with its population counterpart, except that (1) each feature $\boldsymbol{\phi}_i$ or $\boldsymbol{\psi}_j$ is an $\mathbb{R}^n$ vector comprised of the a feature generated from individual samples in $\mathcal{X}$ or $\mathcal{Y}$, and (2) there are a total of $n$ features of either $\mathcal{X}$ or $\mathcal{Y}$. Therefore, the sample distance covariance (or empirical HSIC) can be interpreted as a *finite* weighted sum of pairwise *sample* correlation between one feature of $X$ within "the bag of features of X," $\{\boldsymbol{\phi}_i\}_{i=1}^n$, and another feature of $Y$ within another bag of features of Y, $\{\boldsymbol{\psi}_j\}_{j=1}^n$. Formally, they are the eigenfunctions of $k(\cdot, \cdot)$ or $l(\cdot, \cdot)$ defined over the empirical distribution of $\{\mathbf{x}_i\}$ and $\{\mathbf{y}_j\}$. We will see in Section 4 that smaller $\lambda_i$ and $\sigma_j$ in the ADC of sample distance covariance also corresponds to complex features $\boldsymbol{\phi}_i$ and $\boldsymbol{\psi}_j$, and thereby the correlations of complex features are associated with smaller weight $\lambda_i \sigma_j$. This is a nice characteristic of distance covariance in general, because the correlation between complex features is much less interpretable than the correlation between simple features, and they tend to be attributed to mere chances. Incorporating the weights appropriately reflect the importance of all pairwise correlation.

### 3.3. *Visualization method developed based on the theoretical results*

Sample ADC formula enables us to interpret DC. First, we may construct distance matrices $\mathbf{D}, \mathbf{R}$ (or kernel matrices $\mathbf{K}, \mathbf{L}$), then calculate the spectral decomposition of $-\frac{1}{2}\mathbf{HDH}$ and $-\frac{1}{2}\mathbf{HRH}$ (or $\mathbf{HKH}$ and $\mathbf{HLH}$) to obtain the eigen systems $\{\lambda_i, \boldsymbol{\phi}_i\}$ and $\{\sigma_i, \boldsymbol{\psi}_i\}$. Based on them, *feature dictionary* and *correlation map* illustrate the latent features and their pairwise correlation.

- Feature dictionary

    Feature dictionary calculates and illustrates $\{\boldsymbol{\phi}_i: i = 1, \ldots, I\}$ and $\{\boldsymbol{\psi}_j: j = 1, \ldots, J\}$, the features of both domains $\mathcal{X}, \mathcal{Y}$, where $I$ and $J$ are prescribed number of leading eigen vectors of $\mathbf{K}$ and $\mathbf{L}$. Take domain $\mathcal{X}$ for example. If $p = 1$, we may draw the scatter plot of $(x_n, \boldsymbol{\phi}_i(n))$ for each feature, where $\boldsymbol{\phi}_i(n)$ is the $n$th element of $\boldsymbol{\phi}_i$. If $p = 2$, scatter plots of the dataset $\{\mathbf{x}_n\}$ may be used, where the colors of the individual data points correspond to the value of $\boldsymbol{\phi}_i(n)$. When $p$ is larger than 2, we may first transform each individual sample of $\mathbf{x}_n$ to a point $(x_1, x_2)$ on $\mathbb{R}^2$ plane using low-dimensional embedding methods like t-sne (Hinton and Roweis 2002), then to use color map to illustrate the shapes of the features.

- Correlation Map

    The correlations maps are $I \times J$ image, where each pixel illustrates the intensity of correlation between $\boldsymbol{\phi}_i$ and $\boldsymbol{\psi}_j$, the $i$th feature of $X$ and the $j$th feature of $Y$. Two versions of correlation maps can be used. The *raw correlation map* illustrates the values of $\widehat{\text{corr}}(\boldsymbol{\phi}_i, \boldsymbol{\psi}_j)^2$ at each $(i,j)$ location, which directly shows the correlation between each pair of features. The *weighed correlation map* illustrates the values of $\lambda_i \sigma_j \widehat{\text{corr}}(\boldsymbol{\phi}_i, \boldsymbol{\psi}_j)^2$ with the values of complex features (large $i$ or $j$) tempered, given small values of $\lambda_i$ and $\sigma_j$. The benefit of weighted correlation map is that when all feature correlations are displayed ($I = J = n$), four times the average of all values in the weighted correlation map gives the value of sample DC.

To interpret the results from the feature dictionary and correlation map, we may start with identifying the pairs of features with a large value of $\tilde{C}_{ij}$ that provides the major contribution to the sample DC. Then, we can look up the associated $\boldsymbol{\phi}_i$ and $\boldsymbol{\psi}_j$ from the feature dictionary, to interpret how each feature describe the variability within $X$ and $Y$ respectively. The practitioners may also follow up with scatter plots between $\boldsymbol{\phi}_i$ and $\boldsymbol{\psi}_j$ to examine how the feature $i$ of $X$ and the feature $j$ of $Y$ are related with each other. The visualization method conforms with the DC tests, in the sense that the sum of values in the weighted correlation map is proportional to the sample DC, indicated by the ADC formula. So the ADC-based visualizations give a direct interpretation of the DC results.

### 3.4. *Notes to practitioners*

In engineering practice, the identification of the relationship between the two groups physical variables in signals, spatial data, and images typically follows a two-step procedure. In step 1, a group of features is extracted from each group of variables composing the signals, spatial data and images. Then, the relationship between these two groups of features is investigated through statistical modeling. It is revealed from the ADC formula that the distance covariance method performs a similar procedure. However, there are two unique characteristics of the distance covariance approach that deserves attention from practitioners.

- The latent features $\{\boldsymbol{\phi}_i\}$ and $\{\boldsymbol{\psi}_j\}$ are orthogonal, with a zero mean. They are generated *automatically* from the centered kernel functions $k$ and $l$ through Mercer's theorem, which is ultimately determined by distance metrics $d$ and $\rho$ and the marginal distribution of $\{\mathbf{x}_n\}$ and $\{\mathbf{y}_n\}$. The distinction is that in engineering practices, the features are usually selected based on the meaning of the physical variables, and no orthogonality is guaranteed.
- The distance covariance can be regarded as a composite index calculated from the weighted sum of every pair of features within two groups, where the weight of the correlation coefficient $\lambda_i \sigma_j$ is the product of weights of features in either group. The weights in distance covariance are also automatically generated from the distance metrics $d$ and $\rho$ through the Mercer's theorem, not assigned by practitioners based on their experience and preferences. Although for many kernels, it is rational that the weights with simpler features are associated with larger weights, the weights in the distance covariance cannot be configured arbitrarily and may not match engineering interpretations.

The ADC and visualization method provide valuable information for practitioners to use distance covariance: the feature dictionary enables the practitioners to examine if the automatically generated features match specific engineering interpretations and if the weights match engineer's understanding of the features' importance. If the generated features are not suitable, the practitioner may switch to a different kernel function or distance semi-metric. If the distance covariance test decides that the two variables are dependent, the correlation map will further indicate to which pairs of features the decision is attributed to.

## 4. Experiments

In this section, we conduct a few experiments to illustrates the ADC and the visualization method. The major goal is to demonstrate the application of the feature dictionary correlation map to interpret the correlations. At the same time, we verify the covariance formula, and investigate the difference between multiple kernels or equivalently distance semi-metrics. In the first experiment, we will first study several examples where both $X$ and $Y$ are one-dimensional. In the second experiment, we study a case where $X$ and $Y$ are two-dimensional.

### 4.1. Experiments with one-dimensional X and Y

We interpret the independence testing results from six data sets (Newton 2009, Sarkar and Ghosh 2018), where both $X$ and $Y$ are one-dimensional, as illustrated in Figure 1. Each case contains 500 data points. Below each figure, the two numbers indicate the values of the Pearson correlation coefficient and distance correlation coefficient. In each case, we tried the distance covariance testing with four configurations of distance semi-metrics or kernels of the domain of $\mathcal{X}$ and $\mathcal{Y}$, where Euclidian distance metric is given as $d(x, x') = |x - x'|$, the polynomial kernel function is given as $k(x, x') = (\beta + (x - x')^2)^{-\alpha}$, and the double exponential kernel is given as $k(x, x') = \exp(-(x - x')^2/\alpha)$. The two kernels are selected because they are universal kernels (Micchelli, et al. 2006) and thus the derived tests are consistent against all dependency alternatives. These configurations are summarized in Table 1.

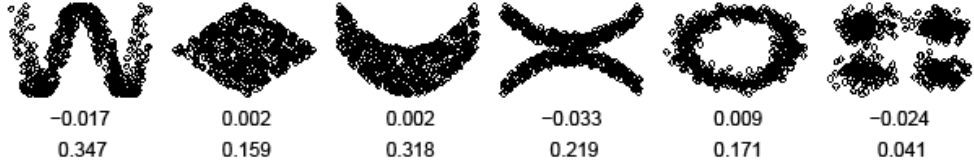

Figure 1 Illustration of six datasets where *X* and *Y* are one-dimensional

Table 1 Four configurations of the distance covariance testing

| Configurations | X | Y |
|---|---|---|
| I | Euclidian distance | Euclidian distance |
| II | Polynomial kernel $\alpha = 0.5, \beta = 0.5$ | Polynomial kernel $\alpha = 0.5, \beta = 0.5$ |
| III | Polynomial kernel $\alpha = 2, \beta = 0.5$ | Polynomial kernel $\alpha = 2, \beta = 0.5$ |
| IV | Double Exponential kernel $\theta = 1$ | Double Exponential kernel $\theta = 1$ |

Newton (2009) validated that distance covariance method is able to reject the independency hypothesis for each of the six cases. In this paper, we will not repeat this hypothesis testing procedure. Instead, we use the ADC formula to interpret and visualize why *X* and *Y* are not independent. Take the data set 1 ("W" shape) for example. Figure 2 illustrates the leading six features ($\phi_i$ and $\psi_j$, $i, j = 1, ..., 6$) generated from the spectral decomposition from both the domains of *X* and *Y* under Configuration I, according to Step 1 of Section 3.3.3. The raw correlation map and the weighted correlation map between each pair of features are illustrated in Figure 3. From the raw correlation map, we can see that the following pairs of features have squared sample correlation coefficients above 0.1: the squared correlation between $\psi_1$ and $\phi_2$, $\phi_4$ are 0.4282 and 0.2582 respectively, and the squared correlation between $\psi_2$ and $\phi_6$ is 0.2497. By checking the feature dictionary, we can know why and *Y* $\psi_1$ and $\phi_2$ are deemed correlated: samples with large values of $\phi_2$ has extreme values in *X* and are distributed at two tails of "W", where *Y* also tends to be large. Thus, it also has large value of $\psi_1$. Similarly, samples with large values of $\phi_4$ are located either at the center, or at the two tails of "W", also correspond to large values of $\psi_1$. These features in the data $(x_i, y_i)$ leads to the large value of $\widehat{\text{corr}}(\phi_2, \psi_1)^2$, thus large value of distance covariance, and thus the rejection of the independent hypothesis. Nevertheless, the features become complex when $i$ or $j$ increases, and therefore interpreting the correlation between $\phi_6$ and $\psi_2$ is hard. In the weighted correlation map, we can see that $\lambda_2 \sigma_1 \widehat{\text{corr}}(\phi_2, \psi_1)^2$ has a highest value of 719.5, $\lambda_4 \sigma_1 \widehat{\text{corr}}(\phi_4, \psi_1)^2 = 114.3$, whereas $\lambda_6 \sigma_2 \widehat{\text{corr}}(\phi_6, \psi_2)^2$ only has a value of 9.7. The total value of $\sum_i \sum_j \lambda_i \sigma_j \widehat{\text{corr}}(\phi_i, \psi_j)^2$ is 868.3, resulting in a distance covariance of $4 \times 868.3/500^2 = 0.0139$. The correlation components of the above three pairs of features take accounts of 97.15% of the total distance covariance. Also, we observe that the correlation of complex features has significantly lower weights: for example, the share of $\lambda_6 \sigma_2 \widehat{\text{corr}}(\phi_6, \psi_2)^2$ is much lower than the other two pairs of features due to the small weight $\lambda_6 \sigma_2$, where $\lambda_6 = 2.82$ only takes 1.68% of the $\sum_i \lambda_i$ and $\sigma_2 = 13.85$ only takes 13.84% of $\sum_j \sigma_j$. The low weights $\lambda_i \sigma_j$ for large $i$ and $j$ can be regarded as an automatic penalization for the correlation of complex features, for which the practitioners have stronger belief that large correlation is driven by chance.

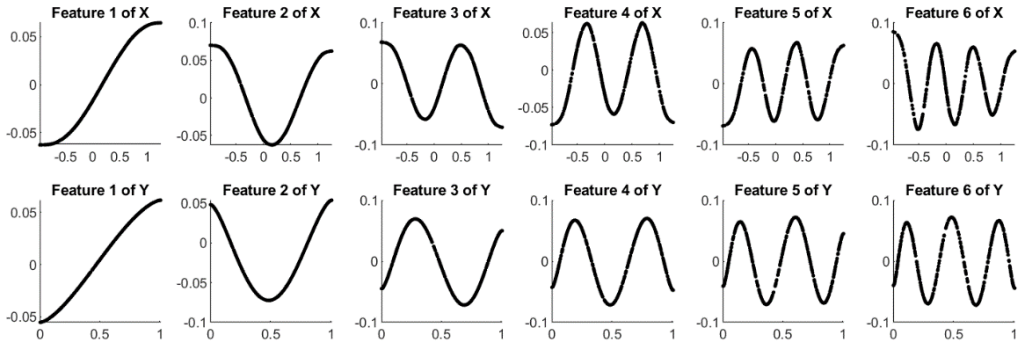

Figure 2 The leading features for $X$ and $Y$ for data set 1 when using Euclidean distance semi-metric

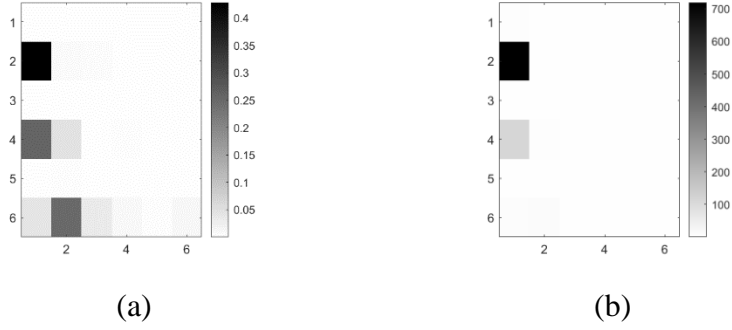

(a)                                              (b)

Figure 3 (a) The raw correlation map of $\widehat{\text{corr}}(\boldsymbol{\phi}_i, \boldsymbol{\psi}_j)^2$, where the $y$ axis denotes $\boldsymbol{\phi}_i$ and $x$ axis denotes $\boldsymbol{\psi}_j$; (b) the squared correlation map $\lambda_i \sigma_j \widehat{\text{corr}}(\boldsymbol{\phi}_i, \boldsymbol{\psi}_j)^2$ in the distance covariance

There are several common observations from the dictionary of features and correlation maps of all six data sets. These results are provided in the supplementary materials.

- First, $\lambda_i$ and $\sigma_j$ decrease rapidly when $i$ and $j$ increases, as shown in Figure 4 (a), which clearly shows that the weights of complex features decrease rapidly as additive components in the distance covariance statistic. Therefore, if $X$ and $Y$ are only dependent through complex features (i.e., $\text{corr}(\phi_i(X), \psi_j(Y)) = 0$ for small $i,j$'s while $\text{corr}(\phi_i(X), \psi_j(Y)) > 0$ for large $i,j$'s), the population distance correlation will be a small positive number and therefore significant amount of samples are needed to reject the independence hypothesis.

- Second, only a small number of weighted correlations between the features contribute to significant percentage of the distance covariance statistics. As we see in Figure 3 (b), only a limited number of blocks are shaded. Specifically, Figure 4 (b) shows how limited pairs of weighted correlation component take a major proportion of total distance covariance. The $x$-axis is the number of pairs of $\lambda_i \sigma_j \widehat{\text{corr}}(\boldsymbol{\phi}_i, \boldsymbol{\psi}_j)^2$, and $y$ axis shows the proportion to the total distance covariance that *cannot* be covered by these pairs of weighted squared correlations. We can see that limited pairs of weighted squared correlations dominate in the ADC Equation (7).

- Third, both the features and weights are dependent on the distribution of $X$ or $Y$ and the kernels or distance semi-metrics used. First, Figure 5 and Figure 6 compares the features generated from dataset 1 (W shape) using polynomial kernel with $\alpha = 0.5, \beta = 0.5$ and with $\alpha = 2, \beta = 0.5$, as well as their weights. We can see that the features generated from two kernels are slightly different (for example, the two ends of feature 1 of $X$). More importantly, with the kernel with $\alpha = 0.5$, the weights for complex features decay more rapidly compared with the kernel with $\alpha = 2$. It indicates that the kernel with $\alpha = 0.5$ puts more testing power on the linear dependency. Second, even when the same kernel function or distance semi-metric is used, the six datasets are associated with slightly different features of $X$ and $Y$ with different weights. It indicates that the latent features and their weights also rely on the marginal empirical distribution of data. This observation implies that the visualization should be performed on a case-by-case basis, for each dataset and the selection of kernels or distance semi-metrics.

## *4.2. Experiments with multi-dimensional X or Y*

In this section, we consider two cases where $X$ and $Y$ are both two-dimensional, where the raw data is illustrated in Figure 7. Colors are added to the data points to illustrate how data points of $X$ and $Y$ are associated with the same sample.

<u>Case I</u>: $X = (Z \cos 2\pi U, Z \sin 2\pi U); Y = (R \cos \pi\theta Z, R \sin \pi\theta Z)$ where $R, U, Z, \theta$ are independent, $R \sim U(0,1)$, $U \sim U(0,1)$. $Z \sim U(0,1)$, and $\theta = \pm 1$ with probability 0.5 each.

<u>Case II</u>: $X = (Z \cos 2\pi U, Z \sin 2\pi U)$, $Y = (\|X - P_1\| + \|X - P_3\| - \|X - P_0\|, \|X - P_2\| + \|X - P_4\| - \|X - P_0\|)$, where $U \sim U(0,1)$. $Z \sim U(0,1)$. The coordinates fo the five points are $P_1(-1,0), P_2(1,0), P_3(0,0.5), P_4(0,-0.5)$ and $P_0(0,0)$.

In the first case, the latent variable $Z$ is associated with the distance to origin for $X$ and the angle for $Y$. It aims to demonstrate a typical situation where the dependency between $X$ and $Y$ is driven by common latent variables, so that the joint distribution of $X$ and $Y$ can be represented as $P_{XY}(x,y) = \int p(x|z)p(y|z)p(z)dz$. In the second case study, our goal is to evaluate our visualization method when the distributions of $X$ and $Y$ are different. For each case, we generated a dataset with the sample size of $N = 1000$. We applied distance covariance with Euclidian distance metrics in both the domain of $X$ and $Y$ to test the independency between them.

We use ADC to visualize the relationship between $X$ and $Y$ for each case. The results of the feature map, the raw correlation map, and the weighted correlation map for each case are illustrated in

Figure 8 and Figure 9, respectively. Distance covariance test results in a close-to-zero p-value for either case, and we aim to use the visualization approach to learn how $X$ and $Y$ are related.

The visualization for case study 1 is given in Figure 9. We can see that the feature $\boldsymbol{\phi}_3$ and the feature $\boldsymbol{\psi}_2$ are most correlated. From the feature dictionary, we can see that feature 3 of $X$ can be interpreted as the norm of each data points, and feature 2 of $Y$ can be interpreted as the horizontal location of the point of the data point. It validates the mechanism of how the data are generated.

Figure 10 provides the visualization for case 2. Multiple pairs of features $\boldsymbol{\phi}_i$, $\boldsymbol{\psi}_j$ are correlated from the raw correlation graph, among which the simplest features $\boldsymbol{\phi}_1, \boldsymbol{\psi}_1$ has the largest weight thus contribute the most to the distance covariance. The points at the bottom right region in the scatter plot of X tends to have large values in feature 1, $\boldsymbol{\phi}_1$, of X, shown the feature dictionary of Figure 10. The points in this region correspond to small $Y_1$ and large $Y_2$, as their distance to $P_1(-1,0)$ and $P_3(0,.5)$ are significantly smaller than their distance to $P_2(1,0)$ and $P_4(0,-.5)$. Therefore, they have a small value of features of $\boldsymbol{\psi}_1$, according to the feature map.

In summary, the experiments above all illustrate that the visualization method can provide extra insights into how the datasets between X and Y are dependent, after a significant dependency relationship between X and Y is identified from the distance covariance method.

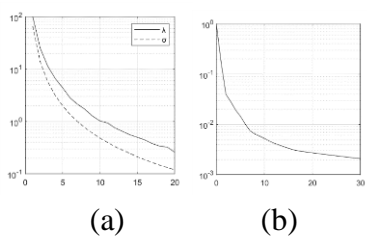

(a)  (b)

Figure 4 (a) The leading features' weight $\lambda_i, \sigma_j$ of dataset 1 when using Euclidean distance semi-metric. (b) The proportion of the remaining components in Equation (7) apart from the largest few pairs of components, indicated with $x$ axis.

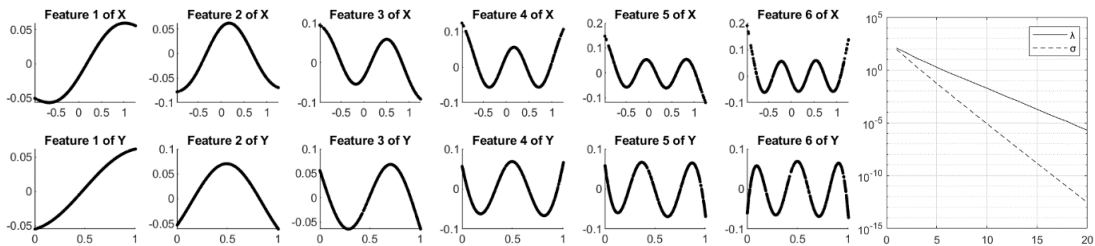

Figure 5 The features and weights from dataset 1 using polynomial kernel with $\alpha = \beta = 0.5$.

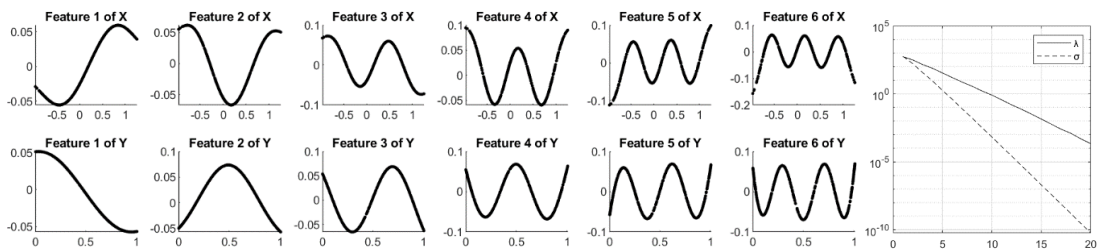

Figure 6 The features and weights from dataset 1 using polynomial kernel with $\alpha = 2, \beta = 0.5$.

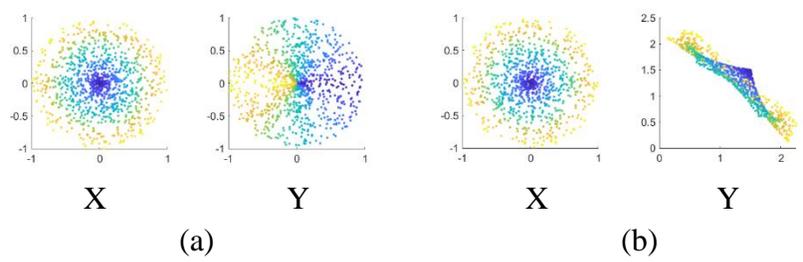

X Y X Y

(a) (b)

Figure 7 Two cases where two dimensional vectors *X* and *Y* are dependent.

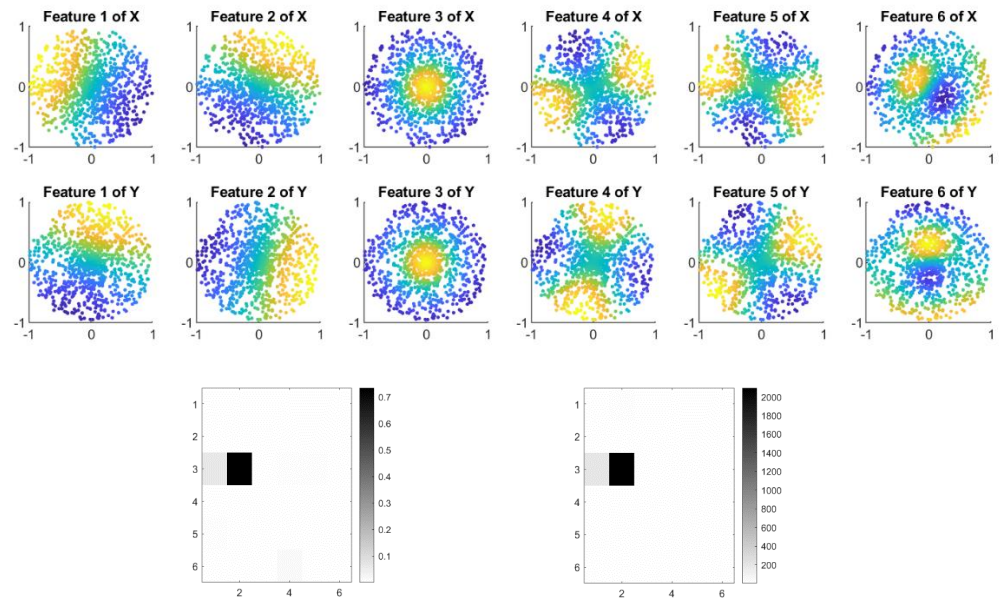

Figure 8 The feature dictionary, the raw correlation map and the weighed correlation map for two-dimensional *X, Y*, case 1.

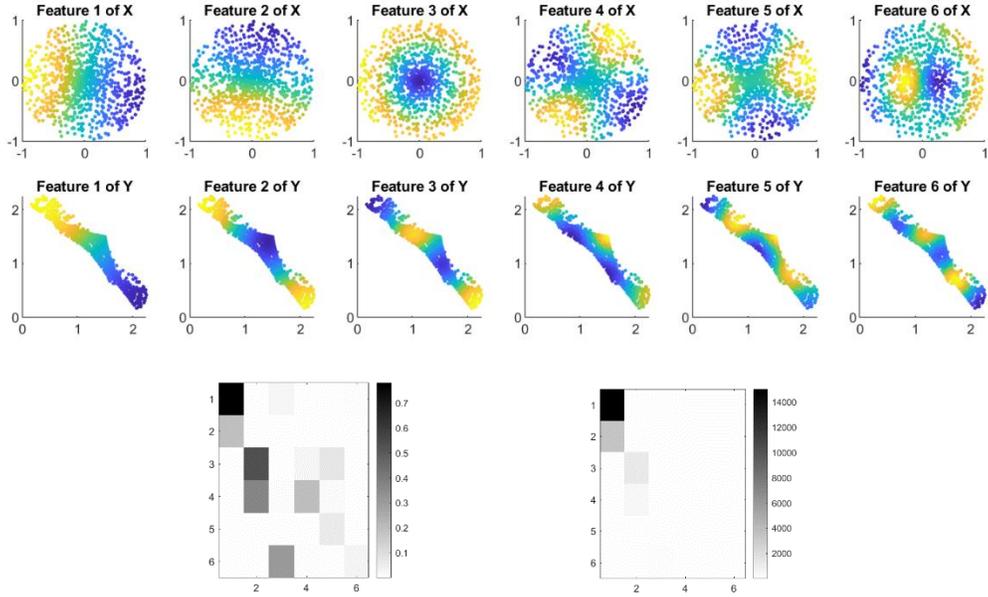

Figure 9 The feature dictionary, the raw correlation map and the weighed correlation map for two-dimensional $X, Y$, case 2.

## 5. Case study

In smart manufacturing, it is important to decide whether there are connections between the sensor measurements obtained from the equipment of the manufacturing process and the quality evaluation of the final product, to facilitate in-process quality control through monitoring the process measurements. It is also necessary to understand how the process data are related to the quality measurements, for both obtaining further assurance on the data-driven decisions and revealing the root cause of the quality issues.

In this case study, we consider the process and quality data obtained from the epitaxy stage of a solar cell manufacturing process, where semiconductor material is deposited on top of substrates through a chemical vapor decomposition. For this manufacturing process, the essential process measurements are the time series of the temperature within the chamber and the surface reflectance. A total of $p = 24$ variables that summarize the characteristics of these time series are obtained from $n = 50$ sample products. After the manufacturing process, the solar conversion efficiency (SCE) $Y$ are measured for these samples. In the end, a matrix of process variables $\mathbf{X} \in$ $\mathbb{R}^{50\times 24}$ and a vector of the quality variables $\mathbf{Y} \in \mathbb{R}^{50\times 1}$ are obtained. The readers may refer to Du, et al. (2018) for a detailed interpretation of these variables.

The distance covariance with the Euclidean distance is applied for testing the dependence between $X$ and $Y$. The p-value obtained from the permutation test is 0.002. We now use the ADC to understand and visualize how $X$ and $Y$ are related.

We start with visualizing the correlation maps. The raw correlation map is shown in Figure 10. It shows that $\boldsymbol{\psi}_1$ is related with $\boldsymbol{\phi}_1$, $\boldsymbol{\phi}_2$ and $\boldsymbol{\phi}_3$, where the sample correlations are similar levels. However, the correlation between $\boldsymbol{\phi}_1$ and $\boldsymbol{\psi}_1$ contributes the most to the distance covariance as can be seen in the weighted correlation map.

After identifying the related features, we aim at understanding what these features represent. In the domain of $Y$, the feature $\boldsymbol{\psi}_1$ is monotonic with the $Y$ value. We compared the empirical distribution of $Y$ for these two groups of data, as in Figure 11. The gap indicates that the samples with positive or negative $\phi_1$ are associated different distributions of $Y$. More specifically, samples with $\phi_1 < 0$ are associated with smaller value of $Y$, thereby resulting in lower value of SCE. To understand $\boldsymbol{\phi}_1$, we observe a pair-wise scatter plots of $X_1, \ldots, X_{24}$ where the values of $\phi_1$ for individual samples are presented by their colors, as shown in Figure 12. The plots in the diagonals of the figure, with gray backgrounds, are the scatter plot between each $X_i$ and the value of $\phi_1$. The figure indicates that $\phi_1$ is significantly correlated with $X_1, X_2, X_3, X_7, X_8, X_9$, where these variables are strongly correlated with each other as well. Therefore, *$\phi_1$ can be regarded as a composite index that reflects the common information of all these correlated features*. To visualize how $\phi_1$ relates to $Y$, we divide all samples into two groups based on $\phi_1 > 0$ or $\phi_1 < 0$. In conclusion, a linear composite index of $X_1, X_2, X_3, X_7, X_8, X_9$ is dependent with the $Y$ value.

In this case study, we are lucky to relate the features $\phi_1, \psi_1$ to specific variables using the visualization method. In general, however, we would point out that there is no guarantee that the features automatically generated from the data allow simple engineering interpretations despite the visualization output, especially when **X** and **Y** are in high dimension. This problem roots in the fact that each feature's values of individual samples are determined directly through the eigen vectors of kernel gram matrix. Therefore, it is hard for us to *customize* the features based on the engineering need. We will leave this problem for future studies.

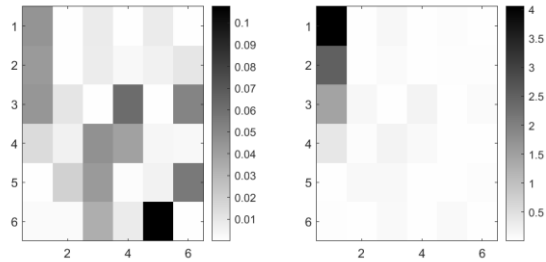

Figure 10 The raw correlation map and the weighted correlation map for the case study.

## 6. Conclusion

Distance covariance and HSIC are known for their power to detect any dependency relationship between two groups of variables. Despite their superiority in testing the dependency, a significant drawback of implementing distance covariance in engineering applications is that the testing results do not directly characterize the relationship between the two groups of variables. This drawback hinders the application of distance covariance by domain experts who could potentially benefit from the result, and also separate the identification of association with the follow-up relationship modeling or diagnostic step. The literature did not provide intuitive interpretations of distance covariance for the practitioners: they are either based on distances of characteristic functions or based on the Hilbert-Schmidt norm of the cross-covariance operators between two RKHSs. The gap between the literature and the real needs from industry and applications motivates us to give an elementary interpretation of distance covariance and develop a visualization method for deciphering the distance covariance testing result.

This study derived an additive decomposition of correlations (ADC) formula for both population and sample distance covariance. This formula leads to a very intuitive interpretation of the distance covariance approach with arbitrary semi-metric: (1) the semi-metrics implicitly generate a class of weighted orthonormal features of both $X$ and $Y$, (2) the distance covariance evaluates the weighted sum of all correlations between individual features.

We designed a visualization method to understand the testing results of distance covariance. The visualization method is based on the ADC formula, and they help to identify how $X$ and $Y$ are dependent. The method is illustrated on multiple simulated and real datasets on revealing how the two groups of variables are independent with each other.

From this paper, the authors hope that the ADC formula clarifies the mechanism of distance covariance to a wider range of applicants and the visualization method provide additional information for revealing dependence between variables. We also hope the insights gained from this mechanism motivate new relationship mining and evaluation approaches for data with more complex structures.

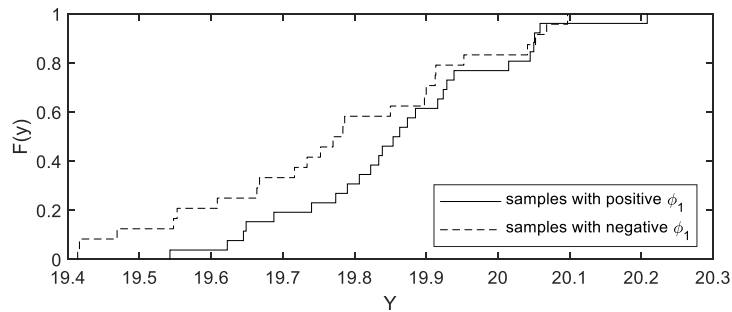

Figure 11 The empirical distribution of $Y$ when $\phi_1$ is positive or negative.

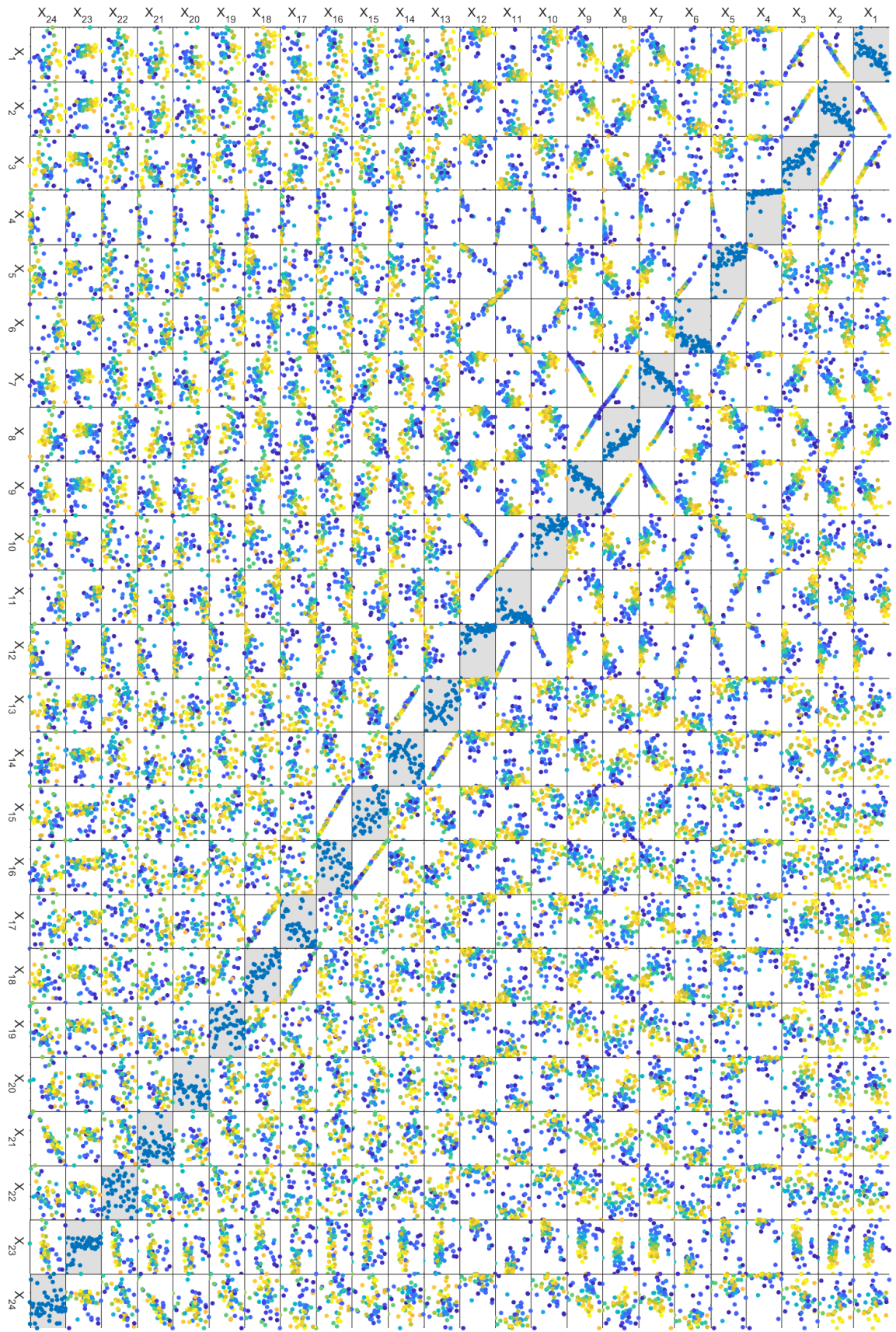

Figure 12 The pairwise matrix plot of $X_1, \ldots, X_{24}$ with the color representing the value of $\phi_1$.

# Appendices

## *Appendix A: Proof of Proposition 1*

For kernel $k$ defined on $(\mathcal{X}, P_X)$, Mercer theorem (Sun 2005) states that

$$k(x, x') = \sum_{i=1}^{\infty} \lambda_i \phi_i(x) \phi_i(x')$$

where the eigen functions $\phi_i \in L_2(\mathcal{X}, \mu)$ with $\int \phi_i(x)\phi_j(x)\, d\mu = \delta_{ij}$, the nonnegative eigenvalues $\{\lambda_i\}$ are absolutely summable, and the series converges absolutely and uniformly $\mu^2$ almost everywhere.

Because the kernels $k(x, x')$ and $l(y, y')$ are centered around $P_X$ and $P_Y$ respectively, we have $\mathbb{E}_{X \sim P_X}[k(X, x')] = 0$, and thereby

$$\mathbb{E}_{X \sim P_X}[\phi_i(X)] = \lambda_i^{-1} \mathbb{E}_{X \sim P_X} \mathbb{E}_{X' \sim P_X}[k(X', X)\phi_i(X')] = \lambda_i^{-1} \mathbb{E}_{X' \sim P_X}\left[\phi_i(X') \mathbb{E}_{X \sim P_X} k(X, X')\right] = 0$$

By the eigen decomposition, $\mathbb{E}_{X \sim P_X}[\phi_i^2(X)] = 1$. Therefore, we know that all features $\phi_i(X)$ have zero mean and unit variance. The same reasoning applies to $\psi_j(Y)$s, and they also have zero mean and unit variance. Put the expression of $k(x, x')$ and $l(y, y')$ to $\text{HSIC}(P_{XY}, k, l)$,

$\text{HSIC}(P_{XY}, k, l)$

$$= \mathbb{E}_{X,X',Y,Y'}[k(X, X')l(Y, Y')] + \mathbb{E}_{X,X'}[k(X, X')]\mathbb{E}_{y,y'}[l(Y, Y')]$$

$$- 2\mathbb{E}_{X,y}\left[\mathbb{E}_{X'}[k(X, X')]\mathbb{E}_{Y'}[l(Y, Y')]\right]$$

$$= \mathbb{E}_{X,X',Y,Y'}\left[\sum_{i=1}^{\infty}\lambda_i\phi_i(X)\phi_i(X')\sum_{j=1}^{\infty}\sigma_j\psi_j(Y)\psi_j(Y')\right]$$

$$+ \mathbb{E}_{X,X'}\left[\sum_{i=1}^{\infty}\lambda_i\phi_i(X)\phi_i(X')\right]\mathbb{E}_{Y,Y'}\left[\sum_{j=1}^{\infty}\sigma_j\psi_j(Y)\psi_j(Y')\right]$$

$$- 2\mathbb{E}_{X,Y}\left[\mathbb{E}_{X'}\left[\sum_{i=1}^{\infty}\lambda_i\phi_i(X)\phi_i(X')\right]\mathbb{E}_{Y'}\left[\sum_{j=1}^{\infty}\sigma_j\psi_j(Y)\psi_j(Y')\right]\right]$$

$$= \mathbb{E}_{X,X',Y,Y'}\left[\sum_{i=1}^{\infty}\sum_{j=1}^{\infty}\sigma_j\lambda_i\phi_i(X)\phi_i(X')\psi_j(Y)\psi_j(Y')\right] + 0 - 0$$

$$= \sum_{i=1}^{\infty}\sum_{j=1}^{\infty}\sigma_j\lambda_i\mathbb{E}_{X,Y}[\phi_i(X)\psi_j(Y)]\mathbb{E}_{X',Y'}[\phi_i(X')\psi_j(Y')]$$

$$= \sum_{i=1}^{\infty}\sum_{j=1}^{\infty}\sigma_j\lambda_i(\text{corr}[\phi_i(X)\psi_j(Y)])^2.$$

The absolute and uniform convergence statement from Mercer's theorem ensures the interchangeability between expectations and summation.

*Appendix B: Proof of Proposition 2*

From $\mathbf{D} = [k(\mathbf{x}_i, \mathbf{x}_i) + k(\mathbf{x}_{i'}, \mathbf{x}_{i'}) - 2k(\mathbf{x}_i, \mathbf{x}_{i'})]_{n\times n}$ and $\mathbf{R} = [l(\mathbf{y}_j, \mathbf{y}_j) + l(\mathbf{y}_{j'}, \mathbf{y}_{j'}) - 2l(\mathbf{y}_{j'}, \mathbf{y}_j)]_{n\times n}$, we can see $\mathbf{HDH} = -2\mathbf{HKH}$ and $\mathbf{HRH} = -2\mathbf{HLH}$, where $\mathbf{H} = \mathbf{I} - \frac{1}{n}\mathbf{J}$. On the other hand,

$$\widetilde{\mathbf{D}} = \widetilde{\mathbf{D}} = \mathbf{D} - \frac{1}{n}\mathbf{DJ} - \frac{1}{n}\mathbf{JD} + \frac{1}{n^2}\mathbf{JDJ} = \mathbf{HDH}, \widetilde{\mathbf{R}} = \mathbf{R} - \frac{1}{n}\mathbf{RJ} - \frac{1}{n}\mathbf{JR} + \frac{1}{n^2}\mathbf{JRJ} = \mathbf{HRH},$$

so $\hat{V}(\mathcal{D}, d, \rho) = \frac{1}{n^2}\text{tr}(\widetilde{\mathbf{D}}^\top\widetilde{\mathbf{R}}) = \frac{1}{n^2}tr(\mathbf{HDHHRH}) = 4\frac{1}{n^2}\text{tr}(\mathbf{HKHHLH}) = \frac{4}{n^2}\text{tr}(\mathbf{KHLH}) = 4\cdot\widehat{\text{HSIC}}(\mathcal{D}; k, l)$, where we used $\mathbf{HH} = \mathbf{H}$.

*Appendix C: Proof of Proposition 3*

Let the spectral decomposition of $\mathbf{HKH} = \mathbf{\Phi\Lambda\Phi}^\top$ and $\mathbf{HLH} = \mathbf{\Psi\Sigma\Psi}^\top$. Put it into the expression of sample HSIC:

$$\widehat{\mathrm{HSIC}}(\mathcal{D}; k, l) = \frac{1}{n^2} \mathrm{tr}(\mathbf{KHLH})$$

$$= \frac{1}{n^2} \mathrm{tr}(\mathbf{\Phi\Lambda\Phi}^\top \mathbf{\Psi\Sigma\Psi}^\top) = \frac{1}{n^2} \mathrm{tr}(\mathbf{\Lambda\Phi}^\top \mathbf{\Psi\Sigma\Psi}^\top \mathbf{\Phi}) = \frac{1}{n^2} \langle \mathbf{\Lambda\Phi}^\top \mathbf{\Psi}, \mathbf{\Phi}^\top \mathbf{\Psi\Sigma} \rangle_F$$

$$= \frac{1}{n^2} \sum_{i=1}^{n} \sum_{j=1}^{n} [\mathbf{\Lambda\Phi}^\top \mathbf{\Psi}]_{ij} [\mathbf{\Phi}^\top \mathbf{\Psi\Sigma}]_{ij} = \frac{1}{n^2} \sum_{i=1}^{n} \sum_{j=1}^{n} (\lambda_i \boldsymbol{\phi}_i^\top \boldsymbol{\psi}_j)(\sigma_j \boldsymbol{\phi}_i^\top \boldsymbol{\psi}_j)$$

$$= \frac{1}{n^2} \sum_{i=1}^{n} \sum_{j=1}^{n} (\lambda_i \sigma_j)(\boldsymbol{\phi}_i^\top \boldsymbol{\psi}_j)^2.$$

Here, vectors $\boldsymbol{\phi}_i$'s and $\boldsymbol{\psi}_j$'s are the columns of $\mathbf{\Phi}$ and $\mathbf{\Psi}$ respectively. From $\mathbf{HKH1} = \mathbf{0}$ and $\mathbf{HLH1} = \mathbf{0}$, we have $\mathbf{\Phi}^\top \mathbf{1} = 0$ and $\mathbf{\Psi}^\top \mathbf{1} = 0$. With $\boldsymbol{\phi}_i^\top \boldsymbol{\phi}_i = 1$ and $\boldsymbol{\psi}_j^\top \boldsymbol{\psi}_j = 1$, we have $\boldsymbol{\phi}_i^\top \boldsymbol{\psi}_j = \widehat{\mathrm{corr}}(\boldsymbol{\phi}_i, \boldsymbol{\psi}_j)$.